\documentclass[final,3p,11pt,authoryear,fleqn]{elsarticle}

\usepackage{float}
\usepackage[none]{hyphenat}
\usepackage{amsmath,amsfonts,amssymb,amsthm,pxfonts,dsfont}
\usepackage{natbib}
\usepackage[bookmarks, colorlinks, pdftitle={Accurate Computation of Marginal Data Density Using Variational Bayes},pdfauthor={Reza Hajargasht and Prasada Rao}]{hyperref}
\hypersetup{linkcolor=blue,citecolor=blue,filecolor=black,urlcolor=blue} 
\usepackage{graphicx} 	
\usepackage{geometry}
\usepackage{multirow}
\usepackage{booktabs}
\usepackage{longtable,lscape,ltcaption} 
\usepackage{rotating} 
\usepackage{setspace}
\usepackage{pifont}
\usepackage{xcolor}

\DeclareMathOperator{\argmin}{argmin}

\begin{document}
%\journal{Something}
%\biboptions{longnamesfirst}
%\theoremstyle{remark}
\newtheorem{thm}{Theorem}
\newtheorem{cl}{Corollary}
\newtheorem{prop}{Proposition}
\newtheorem{lm}{Lemma}
\newdefinition{df}{Definition}
\newdefinition{ex}{Example}
\newdefinition{as}{Assumption}
\newdefinition{pr}{Property}
\newdefinition{rs}{Restriction} 
\newdefinition{al}{Algorithm}
\newproof{pf}{Proof}

\begin{frontmatter}

\title{Multilateral Index Number Systems for International Price Comparisons:\\ 
Properties, Existence and Uniqueness
}%\tnoteref{t2}}
%\tnotetext[t2]{}

\author[sw]{Gholamreza Hajargasht\corref{cor1}}
%\ead{rhajargasht@swin.edu.au}
%\ead[url]{http://www.reza-hajar.com/}

\author[um]{D.S. Prasada Rao\corref{cor2}}
%\ead{d.rao@uq.edu.au}

\cortext[cor1]{Corresponding author. address: 
Swinburne Business School, BA Building, 27 John St, Hawthorn Victoria 3122, Australia, email: \url{rhajargasht@swin.edu.au}}
\address[sw]{Swinburne Business School, Swinburne University of Technology}
\address[um]{School of Economics, University of Queensland}
\cortext[cor2]{The authors wish to thank Erwin Diewert for his helpful comments on an earlier version of this paper. We are indebted to Stephane Gaubert for pointing us to the connection between our fourth theorem and the DAD theorem in the mathematics literature. The authors are thankful to an anonymous referee for the incisive comments and helpful suggestions. We wish also to thank Bill Griffiths and Beth Webster whose comments have improved the exposition of the paper. This research is partly supported by ARC Grant DP 170103559. A part of work by Rao was undertaken when he was visiting the Institute of Economic Research, Hitotsubashi University in 2017.}

\begin{abstract}
Over the past five decades a number of multilateral index number systems have been proposed for spatial and cross-country price comparisons. These multilateral indexes are usually expressed as solutions to systems of linear or nonlinear equations. In this paper, we provide general theorems that can be used to establish necessary and sufficient conditions for the  existence and uniqueness of the Geary-Khamis, IDB, Neary and Rao indexes as well as potential new systems including two generalized systems of index numbers. One of our main results is  that the necessary and sufficient conditions for existence and uniqueness of solutions can often be stated in terms of graph-theoretic concepts and a verifiable condition based on observed quantities of commodities.
\end{abstract}

\begin{keyword}
purchasing power parities, international prices, nonlinear Perron-Frobenius problem,
connected graphs, DAD problem, generalized mean\\
\textit{JEL classification:} 
E31, % Bayesian analysis: General
C19 % Hypothesis testing: General
\end{keyword}

\end{frontmatter}
%%%%%%%%%%%%%%%%%%%%%%%%%%%%%%%%%%%%%%%%%%%%%%%%%
\section{Introduction}\label{sec:introduction}

\noindent Purchasing power parity (PPP) is a widely used multilateral index for comparing price levels and macroeconomic aggregates such as gross domestic product (GDP) and its components across countries. PPPs are now regularly compiled as part of the World Bank's International Comparison Program (ICP). The most recently released findings from the ICP are for the year 2011 covering 177 countries of the world (World Bank, 2015). The PPPs from the ICP are used in assessing the size and distribution of the world economy and rankings of economies. For example, the latest ICP report indicates that the United States was the world's largest economy in 2011 followed by China, India and Japan. Results from the ICP also indicate that there has been a significant reduction in global inequality based on PPP-converted per capita income data. ICP results are also used for calculating World Development Indicators, the Human Development Index (HDI), regional and global poverty, and for comparing health and education expenditures across countries.\\
\indent The PPPs within the ICP are obtained by aggregating price data collected from countries using appropriate multilateral index formulas \footnote{Details of the ICP methodology can be found in \cite{rao13a} and \cite{WorldBank13}.}. A variety of multilateral index numbers have been proposed for the purpose of PPP compilation over the last five decades including but not limited to Gini-Elteto-Koves-Szulc (GEKS); Geary-Khamis (GK); generalized GK; Ikl{\'e} (or IDB); Rao; and the Country-Product-Dummy (CPD). \cite{rao13b} and \cite{diewert2013methods} describe the methods currently employed within the ICP. Economic theoretic approaches to multilateral systems have been discussed in \cite{neary2004rationalizing}, \cite{feenstra2009consistent}, and \cite{feenstra2013shrunk}] while \cite {diewert1988test,diewert1999axiomatic} and \cite{balk2008price,balk2009aggregation} provide overviews of the axiomatic or test approach to multilateral index numbers. \cite{hill2000measuring,hill2009comparing} and \cite{hajargasht2018spatial} discuss spatial chaining methods based on minimum spanning trees and \cite{rao2009purchasing} offers a collection of papers that describe the state of the art and advances that have been made. The quest for indexes with better properties is ongoing and new indexes are being proposed e.g. see \cite{hajargasht2010stochastic} and \cite{rao2016stochastic} for a new stochastic approach.\\
\indent A common attribute of the multilateral index number systems used in international comparisons is that the price indexes from these methods are usually solutions to some suitably formulated systems of equations. These systems can be linear \footnote{Linear here means that equations are linear in unknowns or known functions of the unknowns. For example, a system is considered linear when it is linear in $1/PPP_j$.} as is the case with the GK system or nonlinear as is the case with the \cite{rao1990system} system. These systems can be meaningful only if they have solutions which are positive and unique (up to a factor of proportionality \footnote{A constant multiple of a given set of PPPs leaves price comparisons between countries unchanged. Therefore, it is sufficient if PPPs are determined uniquely up to a factor of proportionality.}). Not surprisingly, some efforts have already been put into proving the existence of solutions to these systems [e.g. \cite{rao1971existence,rao1976existence}, \cite{khamis1972new}; \cite{balk1996comparison,balk2009aggregation}; \cite{neary2004rationalizing}]. However, the existing results do not often assume the most general conditions and do not cover some indexes.\\
\indent This paper contributes to the literature on existence and uniqueness of multilateral indexes in several ways. (i) It provides several theorems for existence and uniqueness of solutions to multilateral indexes in their most general forms. We use these theorems to prove viability of many indexes for some of which the results are new (e.g. for "equally weighted GK" and arithmetic index). We also provide both necessary and sufficient conditions for existence and uniqueness of IDB and Rao indexes. (ii) The paper brings together and clarifies the mathematical concepts and tools required for establishing existence and uniqueness of solutions to different types of multilateral index number systems. (iii) It introduces a new commensurablity axiom for changes in the reference currency units; it is shown that this axiom leads to a class of multilateral index number systems based on generalized means of order $\rho$. This general class encompasses most of the known systems including the commonly used systems. (iv) Another insight from our results is that a compatibility condition is often required when defining an index in the sense that the weights in the equations defining world average prices ($P$s) and the weights in equations defining purchasing power parities ($PPP$s) must be compatible. (v) Finally, an important contribution of the paper is to show that in general, existence and uniqueness of the indexes are guaranteed if the observed quantity matrix is connected.\\
\indent The paper is organized as follows: In Section \ref{sec:Notation} we define basic notations and concepts that underpin the multilateral index number systems considered in the paper. Section \ref{sec:PPP_Multi} describes linear and nonlinear systems including several commonly used index number systems. Section \ref{sec:Main} states and proves the main theorems on existence and uniqueness of general classes of multilateral index numbers. These general theorems are in turn used to prove the existence and uniqueness of many of the index numbers currently used in international comparisons. \ref{sec:Toolkit} provides a mathematical toolkit (including various connectedness concepts, nonlinear eigenvalue theorems and their links to each other) that is used to prove the theorems stated in Section \ref{sec:Main}.

\section{Notation and Basic Concepts}\label{sec:Notation}

\noindent Let $p_{ij}$ and $q_{ij}$ represent the price and the quantity of the $i$-th commodity in the $j$-th country respectively where $i=1,...,N$ and $j= 1,...,M$. We assume that prices are strictly positive and quantities are non-negative \footnote{An implication of this is that expenditure on an item in a country is positive if and only if the corresponding quantity is positive.}. We further assume: (i) for each $i$, $q_{ij}$ is positive for at least one $j$ and (ii) for each $j$, $q_{ij}$ is positive for at least one $i$. As additional notation, we let $\mathbf{p}$ and $\mathbf{q}$ represent $(N \times M)$ matrices of prices and quantities of all commodities in all the countries where $\mathbf{p}$ is strictly positive and $\mathbf{q}$ is non-negative where due to (i) and (ii)  each row and column of $\mathbf{q}$ has at least one strictly positive element.

\subsection{The Purchasing Power Parity}
\noindent Let $PPP_j$ denote the purchasing power parity of currency of country $j$ or the general price level in $j$-th country relative to a numeraire country. $PPP_j$ shows the number of currency units of country $j$ that have the same purchasing power as one unit of currency of a reference country. For example, if PPP for currency of India is equal to INR $15.50$ with respect to one US dollar then $15.50$ Indian rupees in India have the same purchasing power as one US dollar in the United States. We note here that $PPP_j$s would naturally be functions of observed price and quantity data. Multilateral price comparisons consist of the matrix of all binary comparisons. A bilateral price comparison between two countries $j$ and $k$, denoted by $P_{jk}$, is given by 
\begin{equation} \label{eq:1}
P_{jk}\,\,=\,\,\frac{PPP_k}{PPP_j}\,\,\,\forall\,j\,\text{and}\,\,k
\end{equation}
The set of all binary price comparisons, $P_{jk}\, (j,k = 1,2,...,M)$, defined in (\ref{eq:1}) are transitive \footnote{Transitivity requires that all the pairwise comparisons are internally consistent and satisfy the condition $P_{jk}=P_{jl}\times P_{lk} \,\text{for all} j, k \, \text{and}\,\,i$. See \cite{balk2008price} for more on transitivity.} and that the price comparisons are unaffected when all the $PPP_j$s are multiplied by a non-zero constant. Note that, for multilateral price comparisons to be meaningful it is necessary that $PPP_j$s are strictly positive and determined uniquely up to a factor of proportionality, so that $P_{jk}$ in (\ref{eq:1}) is unique.

\subsection{The International Average Price}
\noindent A common feature of the index number systems we study in this paper is that these systems also determine international average prices of the commodities included in the comparisons. Let $P_i$ denote the world average price of the $i$-th commodity $(i =1,2,...,N)$. These $P_i$s are typically expressed as a function of the observed price and quantity data as well as the $PPP_j$s of currencies. Intuitively, the international average price of a commodity is an average of prices of the commodity across countries. As prices in different countries are denominated in different currencies, it is necessary to convert these prices into a common currency unit prior to averaging them. This is achieved by converting observed prices using $PPP$s of currencies. $PPP_j$s are also defined in terms of observed price and quantity data as well as international prices. In Section \ref{sec:PPP_Multi}, we describe different methods for averaging prices such as arithmetic, geometric and harmonic averages.
\subsection{Multilateral Index Number System}\label{sec:PPP}
\noindent A multilateral index number system is an interrelated system of equations which expresses the unknown purchasing power parities and international prices as functions of price and quantity data from different countries. In addition to being functions of observed price and quantity data, each $PPP_j$ is typically a function of all the international prices and similarly each $P_i$ is a function of all the unknown $PPP_j$s thus leading to a system of $M+N$ equations in $M$ unknown $PPP_j$s  and  $N$ unknown $P_i$s. A general multilateral system may be specified as a system of equations of the following form:
\begin{subequations} \label{eq:2}
\begin{align}
  P_i=H_{i}^{1}(\mathbf{PPP,p,q})\quad \quad \quad \quad \quad \quad \quad \quad \quad (i=1,.....,N)  \ \\ 
  PPP_j=H_{j}^{2}(\mathbf{P,p,q})\quad \quad \quad \quad \quad \quad \quad \quad \ \ \ (j=1,.....,M) 
\end{align}
\end{subequations}

\noindent where $\mathbf{PPP}$ and $\mathbf{P}$ are, respectively, $M\times 1$ and $N \times 1$ vectors of purchasing power parities and world average prices. The functions $H_i^{1}$ and $H_j^{2}$ are strictly positive and continuous in all the arguments. $H_i^{1}$s are often in the form of some weighted average of $p_{ij}/PPP_j$ over $j$ and $H_j^{2}$s are in the form of some weighted average of $p_{ij}/P_i$. Different index number systems differ in their specification of functional forms for the equations in (\ref{eq:2}). 

\section{Multilateral Index Numbers Systems Used in ICP}\label{sec:PPP_Multi}
\noindent In this section, we present a number of multilateral systems with an emphasis on those which have been used in the International Comparison Program since its beginning in 1968.
\subsection{Geary-Khamis (GK) and Related Systems}
\noindent The GK system was first proposed by \cite{geary1958note} as a method for comparing agricultural outputs across countries and later extended to its general form by \cite{khamis1972new}. The index was adopted as the main aggregation method in the ICP until 1985 \citep[see][for a discussion of the method]{kravis1982world}. The GK system consists of the following system of $M+N$ equations:
\begin{subequations} \label{eq:3}
\begin{align}
\frac{1}{PPP_j}=\sum_{n=1}^{N}{\frac{p_{nj}q_{nj}}{\sum_{n=1}^{N}{p_{nj}q_{nj}}}}\frac{P_n}{p_{nj}}\quad \quad \quad \ \quad \quad \quad \ (j=1,.....,M) \\ 
 {P_i}=\sum\limits_{m=1}^{M}{\frac{q_{im}}{\sum_{m=1}^{M}{q_{im}}}}\frac{p_{im}}{PPP_m}\quad \quad \quad \quad \quad \quad \ \quad (i=1,.....,N)  
\end{align} 
\end{subequations}
Here, $PPP_j$ is defined as a weighted harmonic mean of price relatives. Similarly, the international average price, $P_i$, is defined as a quantity-share weighted average of prices across countries after they are converted into a common currency using $PPP_j$s. Note that Equation (\ref{eq:3}) is linear in $1/PPP_j$s and $P_i$s. The necessary and sufficient conditions for existence and uniqueness of a solution have been derived by \cite{rao1971existence}, \cite{khamis1972new} and \citet{balk1996comparison}. We provide a proof of existence for a generic linear system so that results for specific systems such as GK can be obtained as corollaries.

\cite{cuthbert1999categorisation} proposed the following generalized class of index number systems in order to show that, with respect to satisfying the additivity property, GK is not unique \footnote{For a definition and further discussion of additivity see \cite{balk2008price} pages 244-251. \cite{cuthbert1999categorisation} disproved a conjecture made by Rao indicating that the GK system is the only multilateral system satisfying additivity. \cite{sakuma2009additivity} provided another system that satisfies additivity but differs from the GK index.}
\begin{subequations} \label{eq:4}
\begin{align}
\frac{1}{PPP_j}=\sum_{n=1}^{N}{\frac{p_{nj}q_{nj}}{\sum_{n=1}^{N}{p_{nj}q_{nj}}}}\frac{P_n}{p_{nj}}\quad \quad \quad \quad \ \quad \quad \quad \ (j=1,.....,M) \\ 
 {P_i}=\sum\limits_{m=1}^{M}{\frac{\beta_mq_{im}}{\sum_{m=1}^{M}{\beta_mq_{im}}}}\frac{p_{im}}{PPP_m}\quad \quad \quad \quad \quad \quad \ \quad (i=1,.....,N)  
\end{align} 
\end{subequations}

\noindent where $\beta_j$s are known constants. The GK system in (\ref{eq:3}) is a special case of (\ref{eq:4}) when $\beta_j=\beta$ for all $j=1,2,...,M$. Other choices for $\beta_j$ could lead to the Ikl{\'e} system \citep[e.g.][pp. 207]{balk1996comparison}. \cite{cuthbert1999categorisation} did not investigate the existence of solutions to the generalized GK system in (\ref{eq:4}). Our Theorem-\ref{thm:th1} provides the necessary and sufficient conditions for the existence of a unique positive solution to this system as a special case.

Note that in systems (\ref{eq:3}) and (\ref{eq:4}), each $PPP_j$ is defined as a weighted harmonic mean of ratios  $\{p_{nj}/P_n, n=1,...,N\}$ and each $P_i$ is an arithmetic mean of the ratios $\{p_{im}/PPP_m, m=1,...,M\}$. A more general system based on generalized means of order $\rho$ can be written as
\begin{subequations} \label{eq:5}
\begin{flalign}
\frac{1}{PPP_j}=\left \{\sum_{n=1}^{N}{w_{nj}\big({\frac{p_{nj}}{P_n}}}\big)^\rho \right \}^{1/\rho}\quad \quad \quad \quad \quad \quad \quad \quad \ (j=1,.....,M) \\ 
 {P_i}=\left \{\sum\limits_{m=1}^{M}{\frac{\beta_jq_{im}}{\sum_{m=1}^{M}{\beta_mq_{im}}}}\big(\frac{p_{im}}{PPP_m}\big)^\rho\right \}^{1/\rho}\quad \quad \quad \quad \quad \quad \ \quad (i=1,.....,N)  
\end{flalign} 
\end{subequations}
where $w_{ij}$ represents the expenditure share of commodity $i$ in country $j$ defined as $w_{ij}=\frac{p_{ij}q_{ij}}{\sum_{n=1}^N{p_{nj}q_{nj}}}$. A common feature of the systems in (\ref{eq:3}), (\ref{eq:4}) and (\ref{eq:5}) is that they share the same type of weights. The $PPP_j$ definition uses expenditure share weights whereas the international prices are defined using quantity share weights. It is easy to show that the GK and generalized GK systems defined above are special cases of the new system defined in (\ref{eq:5}). Of particular interest could be the case with  $\rho\rightarrow 0$ where the new system turns into the geometric version of the GK index.
\subsection{IDB, Rao and Related Systems}
\noindent We now turn to multilateral index number systems that make use of  expenditure shares as weights for defining both $PPP$s and $P$s. The following two sets of weights, $w_{ij}$ and weights $w_{ij}^*$ are used\\
\[w_{ij}=\frac{p_{ij}q_{ij}}{\sum_{n=1}^N{p_{nj}q_{nj}}} \quad \quad \quad
w_{ij}^*=\frac{w_{ij}}{\sum_{m=1}^M{w_{im}}}\]

Here, $w_{ij}$ is the expenditure share of commodity $i$ in country $j$, whereas $w_{ij}^*$ is the expenditure share of commodity $i$ in country $j$ relative to the total share of this commodity across all the countries.

\cite{rao1990system} defines a system for international price comparisons as follows\footnote{\cite{rao2005equivalence} has shown that the solution to this system can be obtained as weighted least squares estimates from the country-product dummy (CPD) method. Further details of the CPD method and its links with multilateral systems are discussed in \cite{hajargasht2010stochastic} and \cite{rao2016stochastic}.}
\begin{subequations} \label{eq:6}
\begin{align}
PPP_j=\prod_{n=1}^{N}{\big(\frac{p_{nj}}{P_n}\big)^{w_{nj}}}\quad \quad \quad \quad \ \quad \quad \quad \ (j=1,.....,M) \\ 
 P_i=\prod_{m=1}^{M}{\big(\frac{p_{im}}{PPP_m}\big)^{w_{im}^*}}\quad \quad \quad \quad \quad \quad \ \quad (i=1,.....,N)  
\end{align} 
\end{subequations}

In the system first proposed by \cite{ikle1972new}, and later simplified and clarified by \cite{dikhanov1997sensitivity} and \cite{balk1996comparison}\footnote{We follow \cite{diewert2013methods} and refer to this system as the Ikl{\'e}-Dikhanov-Balk (IDB) system.}, expenditure share weights are used along with harmonic averages as shown in (\ref{eq:7}).
\begin{subequations} \label{eq:7}
\begin{align}
\frac{1}{PPP_j}=\sum_{n=1}^{N}{\big(w_{nj}\,\frac{P_n}{p_{nj}}\big)}\quad \quad \quad \quad \ \quad \quad \quad \ (j=1,.....,M) \\ 
 \frac{1}{P_i}=\sum_{m=1}^{M}{\big(w_{im}^*\frac{PPP_m}{p_{im}}\big)}\quad \quad \quad \quad \quad \quad \ \quad (i=1,.....,N)  
\end{align} 
\end{subequations}

Note that, in the Rao system in (\ref{eq:6}), $PPP$s and $P$s are defined as geometric means of deflated national prices while in the Ikl{\'e}-Balk-Dikhanov (IDB) system in (\ref{eq:7}), harmonic means of the converted national prices are used in a similar manner.
\cite{hajargasht2010stochastic} proposed a similar system of equations but using arithmetic means: 
\begin{subequations} \label{eq:8}
\begin{align}
PPP_j=\sum_{n=1}^{N}{\big(w_{nj}\frac{p_{nj}}{P_n}\big)}\quad \quad \quad \quad \ \quad \quad \quad \ (j=1,.....,M) \\ 
 P_i=\sum_{m=1}^{M}{\big(w_{im}^*\frac{p_{im}}{PPP_m}\big)}\quad \quad \quad \quad \quad \quad \ \quad (i=1,.....,N)  
\end{align} 
\end{subequations}
\cite{hill2000measuring} proposed the "equally weighted GK system" (EWGK) defined below and found that it has better properties than the GK system 
\begin{flalign*}
\dfrac{1}{PPP_j}=\sum_{n=1}^{N}{w_{nj}\,\dfrac{P_n}{p_{nj}}}\quad \quad \quad \quad \quad \quad  \quad \ (j=1,.....,M) \\ 
 P_i=\sum_{m=1}^{M}{w^*_{im}\,\dfrac{p_{im}}{PPP_m}}\quad \quad \quad \quad \quad \quad \ \quad (i=1,.....,N)  
\end{flalign*} 
where the $PPP_j$ equations are the same as those in the GK and Ikl{\'e} indexes and the $P_i$ equations are the same as those in the arithmetic index.  
A new general index number system which encompasses the systems described in (\ref{eq:6}), (\ref{eq:7}) and (\ref{eq:8}) can be defined as:
\begin{subequations} \label{eq:9}
\begin{align}
PPP_j=\left \{\sum_{n=1}^{N}{w_{nj}\big(\frac{p_{nj}}{P_n}\big)}^\rho\right\}^{1/\rho}\quad \quad \quad \quad \ \quad \quad \quad \ (j=1,.....,M) \\ 
 P_i=\left \{\sum_{m=1}^{M}{w_{im}^*\big(\frac{p_{im}}{PPP_m}\big)}^\rho\right\}^{1/\rho}\quad \quad \quad \quad \quad \quad \ \quad (i=1,.....,N)  
\end{align} 
\end{subequations}
Different values for $\rho$  leads to different indexes. For example $\rho=0$ leads to the Rao system, $\rho=-1$ gives the IDB system and $\rho=1$ leads to the arithmetic index \footnote{It might be possible to associate a stochastic model to this system and therefore estimate $\rho$ or statistically test between these indexes.}. Theorem-\ref{thm:th5} proved in this paper can be used to establish the existence and uniqueness of all of these systems and more.
\subsection{\cite{neary2004rationalizing} and \cite{rao1976existence} Systems}  
\noindent The \cite{neary2004rationalizing} and \cite{rao1976existence} systems bring economic theory and the notion of cost of living indexes into the definition of \textit{PPP}s. The Konus cost of living index is defined as the ratio of minimum expenditure required to attain a certain level of utility at two different sets of prices  \footnote{\cite{diewert1976exact} discusses the Konus index in detail.}. The \cite{neary2004rationalizing} system uses the Konus index to define $PPP$s as follows:
\begin{subequations} \label{eq:10}
\begin{align}
\frac{1}{PPP_j}=\sum_{n=1}^{N}{\frac{p_{nj}q_n^*(\mathbf{P,q_j})}{\sum_{n=1}^{N}{p_{nj}q_{nj}}}}\frac{P_n}{p_{nj}}\quad \quad \quad \ \quad \quad \quad \quad \ (j=1,.....,M) \\ 
 {P_i}=\sum\limits_{m=1}^{M}{\frac{q_{im}}{\sum_{m=1}^{M}{q_i^*(\mathbf{P,q_m})}}}\frac{p_{im}}{PPP_m}\quad \quad \quad \quad \quad \quad \ \quad (i=1,.....,N)  
\end{align} 
\end{subequations}

\noindent $q_i^*(\mathbf{P,q_j})$ \footnote{The notation $q_i^*(\mathbf{P,q_j})$ could be replaced by $q_{ij}^*$ to indicate that these quantities depend upon the utility obtained by country $i$ with quantity vector $\mathbf{q}_j$.} in (\ref{eq:10}a) and (\ref{eq:10}b) are optimal (cost-minimizing) quantities obtained as the solution to the following cost minimization problem where $U(.)$ is a well-behaved utility function \footnote{Further details can be found in \cite{neary2004rationalizing} and \cite{rao1976existence}.}
\[\argmin_{(q_1^*\, ,..., \,q_N^*)} \sum_{n=1}^N{P_nq_{n}^*} \quad  \text{subject to} \quad  U(q_1^*,..,q_N^*)\geq U(\mathbf{q_j}) \tag{10c}\]
This problem is solved for each country separately. 
The main difference between the Neary system and the GK system is in the use of unobserved quantities, $q_{ij}^*$. In the Neary system we have quantities $q_{ij}^*$s which depend on the vector of international average prices $\mathbf{P}$. This calls for a different theorem for establishing existence and uniqueness. 

The system proposed by \cite{rao1976existence} is similar to the \cite{neary2004rationalizing} system except that the international average prices are defined differently:
\begin{equation} \label{eq:11}
{P_i}=\sum\limits_{m=1}^{M}{\frac{q_{im}}{\sum_{m=1}^{M}{q_{im}}}}\frac{p_{im}}{PPP_m}\quad \quad \quad \quad \quad \quad \ \quad (i=1,.....,N)  
\end{equation}

\noindent \cite{rao1976existence} studied existence of solutions to this system by using a version of a nonlinear eigenvalue theorem but his proof is incomplete. \footnote{In fact, he could not prove that $\lambda^*=1$ (see Section \ref{sec:Main} for the relevance of the condition) and this is not expected to be the case in general due to lack of a compatibility condition.}   
\section{Existence Theorems for Multilateral Index Number Systems} \label{sec:Main}
\noindent As we mentioned before, multilateral index number systems can be classified into two groups: The  first group are essentially linear systems in terms of $PPP_j$s, or $P_i$s, or in terms of some functions of $PPP_j$s and $P_i$s whereas the second group are nonlinear. In this section we provide theorems that can be applied to both cases. For stating and proving of these theorems we need a variety of mathematical concepts and tools that are summarized in \ref{sec:Toolkit}.
\subsection{An Existence Theorem for Linear Systems}
\noindent The following theorem proves existence and uniqueness for the linear class of systems in its most general form.
\begin{thm} \label{thm:th1} Consider the following general system of $M+N$ equations:
 \begin{subequations} \label{eq:15}
\begin{flalign}
f_j(PPP_j)=\sum_{n=1}^{N}{a_{nj}\,g_n(P_n)}\quad \quad \quad \quad \quad \quad  \quad \ (j=1,.....,M) \\ 
 g_i({P_i})=\sum_{m=1}^{M}{b_{im}\,f_m(PPP_m) }\quad \quad \quad \quad \quad \quad \ \quad (i=1,.....,N)  
\end{flalign} 
\end{subequations}
where $f_j(.)$ and $g_i(.)$ can be any bijective functions from $R_{+}\to R_{+}$; $a_{ij}$ and $b_{ij}$ are non-negative weights. Under the following assumption:\\
(T.1) \, $a_{ij}$ and $b_{ij}$ can be written such that $a_{ij}=\dfrac{d_{ij}}{\sum_{n=1}^{N}{c_{nj}}}$ and $b_{ij}=\dfrac{c_{ij}}{\sum_{m=1}^{M}{d_{im}}}$ with  ${q_{ij}}>0\Leftrightarrow {d_{ij}}>0$ and \hspace*{1cm} ${q_{ij}}>0\Leftrightarrow {c_{ij}}>0$.\\
\noindent A necessary and sufficient condition for the existence of a unique positive solution (up to a positive scalar factor) is \hyperlink {CM} {connectedness} of the quantity matrix $\mathbf{q}$. 
\end{thm}

Before offering a proof for this theorem, we examine some aspects of the theorem.
\begin{enumerate} [\textbullet]
\item Note that for existence of a non-trivial solution, $a_{ij}$ and $b_{ij}$ in equation (\ref{eq:15}) cannot be completely independent and a \textit{compatibility} condition is needed. One such condition is given by Assumption (\textit{T.1}) above. \cite{khamis1989gerardi} considered an interesting system in which conditions $T.1$ is not satisfied; they showed that the system has only a trivial solution.
\item As we discussed in defining PPP indexes, each $PPP_j$ is often set as some kind of average of $p_{ij}$s (over $i$) deflated by international prices $P_i$s and each $P_i$ as some kind of average of $p_{ij}$s (over $j$) deflated by $PPP_j$s. With appropriate definitions for $d_{ij}$ and $c_{ij}$, $f_j$ and $g_i$ we can easily cover such indexes.  As an example, consider the GK system (\ref{eq:3}) i.e. 
\begin{align*}
\frac{1}{PPP_j}=\sum_{n=1}^{N}{\frac{p_{nj}q_{nj}}{\sum_{n=1}^{N}{p_{nj}q_{nj}}}}\frac{P_n}{p_{nj}}\quad \quad \quad \ \quad \quad \quad \ (j=1,.....,M) \\ 
 {P_i}=\sum\limits_{m=1}^{M}{\frac{q_{im}}{\sum_{m=1}^{M}{q_{im}}}}\frac{p_{im}}{PPP_m}\quad \quad \quad \quad \quad \quad \ \quad (i=1,.....,N)  \end{align*} 
It is easy to see that by defining $d_{ij}=q_{ij}$, $c_{ij}=p_{ij}q_{ij}$, $f_j(x)=1/x$ and $g_i(x)=x$ in Theorem-\ref{thm:th1} we obtain this system.  
\end{enumerate}

To prove the theorem, we establish the following two lemmas where we use the notation $f_j=f_j(PPP_j)$ and $g_i=g_i(P_i)$. If we obtain solutions for $f_j$ and $g_i$, then we can obtain solutions for $PPP_j$ and $P_i$ by invoking the bijective nature of these functions.

\begin{lm}\label{lm:lm1} 
Consider the following system of equations defined in terms of $(\mathbf{f, g})$
\begin{subequations} \label{eq:16}
\begin{flalign}
f_j=\sum_{n=1}^{N}{\dfrac{d_{nj}}{\sum_{n=1}^{N}{c_{nj}}}\,g_n}\quad \quad \quad \quad \quad \quad  \quad \ (j=1,.....,M) \\ 
 g_i=\sum_{m=1}^{M}{\dfrac{c_{im}}{\sum_{m=1}^{M}{d_{im}}}\,f_m }\quad \quad \quad \quad \quad \quad \ \quad (i=1,.....,N)  
\end{flalign} 
\end{subequations}
then a necessary and sufficient condition for the existence of a unique positive $\mathbf{f^*}=(f_1^*,...,f_M^*)'$ and $\mathbf{g^*}=(g_1^*,...,g_N^*)'$ (up to a positive scalar factor) is \hyperlink {IR}{irreducibility} of matrices $\mathbf{B, C}$ or $\mathbf{D}$ defined below.
\end{lm}
\noindent \textit{Proof of Necessity:}
Through direct substitution, we first express the system (\ref{eq:16}) in
matrix form\\

\hspace*{1.5cm} $\left[ \begin{matrix}
   0\  & \cdots  & 0 & \dfrac{{{c}_{11}}}{\sum\limits_{n=1}^{N}{{{c}_{n1}}}} & \cdots  & \dfrac{{{c}_{1M}}}{\sum\limits_{n=1}^{N}{{{c}_{nM}}}}  \\
   \vdots  & {} & \vdots  & \vdots  & {} & {}  \\
   0 & \cdots  & 0 & \dfrac{{{c}_{N1}}}{\sum\limits_{n=1}^{N}{{{c}_{n1}}}} & \cdots  & \dfrac{{{c}_{NM}}}{\sum\limits_{n=1}^{N}{{{c}_{nM}}}}  \\
   \dfrac{{{d}_{11}}}{\sum\limits_{m=1}^{M}{{{d}_{1m}}}} & \cdots  & \dfrac{{{d}_{N1}}}{\sum\limits_{m=1}^{M}{{{d}_{Nm}}}} & 0 & \cdots  & 0  \\
   \vdots  & {} & \vdots  & \vdots  & {} & \vdots   \\
   \dfrac{{{d}_{1M}}}{\sum\limits_{m=1}^{M}{{{d}_{1m}}}} & \cdots  & \dfrac{{{d}_{NM}}}{\sum\limits_{m=1}^{M}{{{d}_{Nm}}}} & 0 & \cdots  & 0  \\
\end{matrix} \right]\left[\begin{matrix} {{g}_{1}}\sum\limits_{m=1}^{M}{{{d}_{1m}}} \\ 
   \ \vdots  \\ 
  \ \vdots  \\ 
  {{g}_{N}}\sum\limits_{m=1}^{M}{{{d}_{Nm}}} \\ 
  {{f}_{1}}\sum\limits_{n=1}^{N}{{{c}_{n1}}} \\ 
   \ \vdots  \\ 
  \ \vdots  \\ 
  {{f}_{M}}\sum\limits_{n=1}^{N}{{{c}_{nM}}}\end{matrix} \right]=\left[\begin{matrix}  {{g}_{1}}\sum\limits_{m=1}^{M}{{{d}_{1m}}} \\ 
   \ \vdots  \\ 
  \ \vdots  \\ 

  {{g}_{N}}\sum\limits_{m=1}^{M}{{{d}_{Nm}}} \\ 
  {{f}_{1}}\sum\limits_{m=1}^{N}{{{c}_{m1}}} \\ 
   \ \vdots  \\
  \ \vdots  \\ 

  {{f}_{M}}\sum\limits_{m=1}^{N}{{{c}_{mM}}} \\ 
 \end{matrix} \right]$
 \\\\
 or more stacked form  
  $\left( \begin{matrix}
   \mathbf{0} & \mathbf{C}  \\
   \mathbf{D} & \mathbf{0}  \\
\end{matrix}  \right) \left(\begin{matrix}
  {\mathbf{X}_1} \\ 
  {\mathbf{X}_2} \end{matrix}\\ 
 \right)=\left(\begin{matrix}
  {\mathbf{X}_1} \\ 
  {\mathbf{X}_2} \end{matrix}\\ 
 \right)$ where $\mathbf{C}$, $\mathbf{D}$, $\mathbf{X_1}$ and $\mathbf{X_2}$ can be identified from the above expression and $\mathbf{B}=\left(\begin{matrix}
   \mathbf{0} & \mathbf{C}  \\
   \mathbf{D} & \mathbf{0}  \\
\end{matrix}  \right)$.\\
Note that by defining $\mathbf{F=DC}$, we can focus on the subsystem involving only $\mathbf{f}=(f_1,...,f_M)'$:
\\
\begin{equation} \label{eq:17}
\hspace*{0.5cm} \mathbf{F}\mathbf{X}_2=\mathbf{X}_2\Rightarrow \left[ \begin{matrix}
   \dfrac{\sum\limits_{n=1}^{N}{\dfrac{{{d}_{n1}}{{c}_{n1}}}{\sum\limits_{m=1}^{M}{{{d}_{nm}}}}}}{\sum\limits_{n=1}^{N}{{{c}_{n1}}}} & \cdots & \cdots & \dfrac{\sum\limits_{n=1}^{N}{\dfrac{{{d}_{n1}}{{c}_{nM}}}{\sum\limits_{m=1}^{M}{{{d}_{nm}}}}}}{\sum\limits_{n=1}^{N}{{{c}_{nM}}}}  \\
   \vdots  &  &  & \vdots   \\
    \vdots  &  &  & \vdots   \\
   \dfrac{\sum\limits_{n=1}^{N}{\dfrac{{{d}_{nM}}{{c}_{n1}}}{\sum\limits_{m=1}^{M}{{{d}_{nm}}}}}}{\sum\limits_{n=1}^{N}{{{c}_{n1}}}} & \cdots & \cdots & \dfrac{\sum\limits_{n=1}^{N}{\dfrac{{{d}_{nM}}{{c}_{nM}}}{\sum\limits_{m=1}^{M}{{{d}_{nm}}}}}}{\sum\limits_{n=1}^{N}{{{c}_{nM}}}}  \\
\end{matrix} \right]\left[\begin{matrix}
   {{f}_{1}}\sum\limits_{n=1}^{N}{{{c}_{n1}}} \\ 
  \   \\ 
  \ \vdots  \\ 
  \ \vdots  \\
  \   \\ 
   {{f}_{M}}\sum\limits_{n=1}^{N}{{{c}_{nM}}} \\ \end{matrix} 
 \right]=\left[\begin{matrix}
   {{f}_{1}}\sum\limits_{n=1}^{N}{{{c}_{n1}}} \\ 
  \   \\ 
  \ \vdots  \\ 
  \ \vdots  \\
  \   \\ 
   {{f}_{M}}\sum\limits_{n=1}^{N}{{{c}_{nM}}} \\ \end{matrix} 
\right]
\end{equation}

Alternatively, we can define $\mathbf{E=CD}$ and focus on a subsystem involving only $\mathbf{g}=(g_1,...,g_N)'$. We can write each of these linear systems as $\mathbf{AX=X}$ where matrix $\mathbf{A}$ (i.e. $\mathbf{F}, \mathbf{E}$ or $\mathbf{B}$) is a non-negative matrix with columns that sum to one. To prove the \textit{necessity} part of the theorem, we note that the matrix $\mathbf{A'}$, transpose of $\mathbf{A}$ , is a stochastic matrix \footnote{A matrix is said to be stochastic if it is non-negative and each row sums to one.} (each column of matrix $\mathbf{A'}$ sums to 1) and therefore it has a dominant eigenvalue equal to one with corresponding eigenvector equal to $(1, ... ,1)$ \citep[see e.g.][pp 100]{gantmacher2005applications}. Now, for $\mathbf{AX=X}$ to have a unique positive solution, matrix $\mathbf{A}$ must have a dominant eigenvalue equal to one with a corresponding positive eigenvector. But this (i.e. both $\mathbf{A}$ and $\mathbf{A'}$ having the same dominant eigenvalue with positive eigenvectors) is possible only if $\mathbf{A}$ is irreducible \citep[see corollary in page 96 of][]{gantmacher2005applications}.\\
To prove \textit{sufficiency}, we use the Perron-Frobenius theorem. According to this theorem if matrix $\mathbf{A}$ is irreducible then $\mathbf{AX}=\lambda \mathbf{X}$ has a unique positive solution with $\lambda > 0$. Furthermore, since each column of $\mathbf{A}$ sums to one, we must have $\lambda=1$. Note also that properties of the $\mathbf{q}$ matrix ensure that  $\sum_{j=1}^M{d_{ij}}>0$ and $\sum_{i=1}^N{c_{ij}}>0$ therefore $\mathbf{f^*}=(f_1^*,...,f_M^*)'$  and $\mathbf{g^*}=(g_1^*,...,g_N^*)'$ are well-defined. \footnote{Note that connectedness is necessary for uniqueness of the solutions. Without connectedness there can be solutions but they are not unique since the system of equations for some $J \subset {1,...,M}$ can be divided into at least two independent subsystems with independent solutions $\mathbf{f}_{J}^*=\{f_j^*|j \in J\}$ and $\mathbf{f}_{J^c}^*=\{f_j^*|j \in J^c\}$ with $\gamma_1\mathbf{f}_{J}^*$ and $\gamma_2\mathbf{f}_{J^c}^*$ for any $\gamma_1>0$ and for any $\gamma_2>0$ which violates uniqueness.}
\begin{lm} \label{lm:lm2} 
A necessary and sufficient condition for matrix $\mathbf{F}$ in (\ref{eq:17}), (and therefore $\mathbf{E}$ or $\mathbf{B}$), to be irreducible is the connectedness of the set of countries based on the quantity matrix $\mathbf{q}$.
\end{lm}

\noindent \textit{Proof of Sufficiency}: If $\mathbf{q}$ is connected, then for any non-empty  $J \subset \{1,2,...,M\}$, there exists at least one $k \in J$ and $i \in \{1,...., N\}$ such that $q_{ik}>0$ and at least one $j \notin J$ such that $q_{ij}>0$. Note that this implies $d_{ik}>0, c_{ik}>0, d_{ij}>0, c_{ij}>0$. The \textit{sufficiency} is proved if we show that $\{\mathbf{F}\}_{kj}$ (i.e. element $k,j$ of the matrix $\mathbf{F}$) is positive. Note that\\

\hspace*{2cm}${{\{\mathbf{F}\}}_{kj}}=\dfrac{\sum\limits_{n=1}^{N}{{{{d}_{nk}}{{c}_{nj}}}\,\big/{\sum\limits_{m=1}^{M}{{{d}_{nm}}}}\;}}{\sum\limits_{n=1}^{N}{{{c}_{nj}}}}$
\\\\
It can be seen that this is positive, if and only if, there is at least one $i \in {1,...., N}$ such that both $d_{ik}>0$ and $c_{ij}>0$, but connectedness as argued in the previous paragraph guarantees existence of such an  $i$.\\
\textit{Proof of Necessity}: If matrix $\mathbf{F}$ is irreducible, then for any $J \subset \{1,2,...,M\} \neq \varnothing $  there exists at least one $k \in J$ and $j \notin J$ such that $\{\mathbf{F}\}_{kj}>0$  which implies existence of at least one $i \in {1,...,N}$ such that $d_{ik}>0$ and $c_{ij}>0$  but this in turn implies existence of $q_{ik}>0$ and $q_{ij}>0$ and therefore connectedness. \\

\noindent \textit{Proof of Theorem-\ref{thm:th1}}: From Lemma-\ref{lm:lm1}, irreducibility of matrices $\mathbf{F}$ ($\mathbf{E}$ or $\mathbf{B}$) is necessary and sufficient for the existence of a unique positive solution for the system (\ref{eq:15}) in Theorem-\ref{thm:th1}. Lemma-\ref{lm:lm2} shows that connectedness of the quantity matrix $\mathbf{q}$ is necessary and sufficient for irreducibility of the matrices involved thus establishing Theorem-\ref{thm:th1}. 

We can restate the necessary and sufficient conditions stated in Theorem-\ref{thm:th1} on  the quantity matrix $\mathbf{q}$ as: unique positive solutions to multilateral systems of the form (\ref{eq:15}) exist if and only if the \hyperlink {CA}{adjacent graph $\mathbf{G}_q$ is connected} (see \ref{sec:Toolkit} for further information).

\subsection{Choice of Functional Forms for $\mathbf{f} (.)$ and $\mathbf{g} (.)$}
\noindent Theorem-\ref{thm:th1} stated above is quite general and functions $\mathbf{f}$ and $\mathbf{g}$ are (otherwise unrestricted) positive bijective functions.  In this section, we introduce a simple axiom that underpins meaningful international price comparisons and show that under this axiom the functions, $\mathbf{f}$ and $\mathbf{g}$, must have a particular form.
In the Lemmas and Theorem-\ref{thm:th1} proved above, we have shown that if there is a positive $\mathbf{f^*}=(f_1^*,...,f_M^*)'$  and $\mathbf{g^*}=(g_1^*,...,g_N^*)'$ that solves the system of equations then  $\delta\mathbf{f}^*$ and $\delta\mathbf{g}^*$ for every $\delta>0$ is also a solution. We have assumed that $f_j$ and $g_i$ are invertible functions, therefore there exist vectors $\mathbf{PPP}$ and $\mathbf{P}$ that solve system (\ref{eq:15}). However, in general if $PPP_j$s and $P_i$s are solutions then $f_j^{-1}\{\delta f_j(PPP_j)\}$ and $g_i^{-1}\{\delta g_i(P_i)\}$ are also solutions.

In order to narrow the class of functions $\mathbf{f} (.)$ and $\mathbf{g} (.)$, we invoke the following axiom which states that if the unit of measurement of the reference currency is multiplied by $\gamma$ then the international average price of the commodity must be divided by $\gamma$. For example, if the reference currency is 1 US dollar and if the international average price of wheat per tonne is US \$125; then when the reference currency is changed to a 100 US dollar unit of currency then the international average price would be 1.25 units of reference currency (US \$100 unit).

\noindent \textit{Axiom of Change of Reference Currency Unit}: This axiom simply states that if $\mathbf{PPP^*}$ and $\mathbf{P^*}$ are solutions to the multilateral system, then  $\gamma \mathbf{PPP^*}$ and $(1/\gamma)\mathbf{P^*}$ should also be solutions for every $\gamma>0$ .

This axiom helps us to narrow the class of functions $\mathbf{f}(.)$ and $\mathbf{g}(.)$ that can be used in international comparisons. Theorem-\ref{thm:th2} below establishes the functional forms for $\mathbf{f}$ and $\mathbf{g}$ that satisfy the axiom of units of reference currency unit.

\begin{thm} \label{thm:th2}
The axiom of change of reference currency unit is both necessary and sufficient
for the functions $f_j(.)$ and $g_i(.)$ to be of the form $f_j(x)=\alpha_jx^{\rho}$ and $g_i(x)=\beta_i x^{ -\rho}$  for any $\alpha_j>0$, $\beta_i>0$ and $\rho \in R$. \footnote{Here we exclude trivial situations where either $f_j(x)=0$ or $g_i(y)=0$.}
\end{thm}
\noindent \textit{Proof of Necessity}: Note that having both $[\mathbf{PPP,P}]$ and $[\gamma \mathbf{PPP}, (1/\gamma)\mathbf{P}]$ as solutions for any $\gamma>0$ is equivalent to having both $[f_j(PPP_j), g_i(P_i)]$ and $[f_j(\gamma PPP_j), g_i(1/\gamma)P_i)]$  as solutions to equations in (\ref{eq:15}). On the other hand, according to Theorem-\ref{thm:th1} if $[f_j(PPP_j),g_i(P_i)]$ is a solution then $[\delta f_j(PPP_j),\delta g_i(P_i)]$ is also a solution for every $\delta>0$. Therefore, for every $\gamma>0$ there exists some  $\delta>0$ such that:
\begin{flalign*}
 {{f}_{j}}(\gamma PPP_{j}^{{}})\,=\,\delta {{f}_{j}}(PPP_{j}^{{}})\,\,\,\,\quad \quad \quad \quad \ j=1,2,...,M \\ 
 {{g}_{i}}(\frac{1}{\gamma }P_{i}^{{}})\,\,=\,\delta {{g}_{i}}(P_{i}^{{}})\,\,\,\,\,\,\,\,\,\,\,\,\,\,\quad \quad \quad \quad \ i=1,2,...,N  
 \end{flalign*}
Since we are assuming this to be true for all possible solutions $[\mathbf{PPP,P}]>\mathbf{0}$, we apply these conditions to the case where $PPP_j=1$  and $P_i=1$  (e.g. when all prices in all countries are the same) and have
\begin{flalign*}
 {{f}_{j}}(\gamma)\,=\,\delta {{f}_{j}}(1)\,\,\,\,\quad \quad \quad \quad (j=1,2,...,M) \\ 
 {{g}_{i}}(\frac{1}{\gamma })\,=\,\delta {{g}_{i}}(1)\,\,\,\,\quad \quad \quad \quad (i=1,2,...,N)  
\end{flalign*}
Then the equations can be rewritten as:
\begin{flalign*}
 {{f}_{j}}(\gamma PPP_{j}^{{}})\,=\, \dfrac{f_j({\gamma})}{f_j(1)} {f_j(PPP_j)}\,\,\,\,\quad \quad \quad \quad \text{for}\ j=1,2,...,M \\ 
 {g_i}(\frac{1}{\gamma }P_{i}^{{}})\,\,=\,\ \dfrac{g_i(1/\gamma)}{g_i(1)} {g_i}(P_i)\,\,\,\,\,\,\,\,\,\quad \quad \quad \quad \text{for}\ i=1,2,...,N  
 \end{flalign*}
 where $f_j(\gamma)$ and $g_i(1/\gamma)$ are continuous functions defined over $R_+$. In general, each equation in the above system is a special form of the 4th Pexider’s functional equation \citep[see e.g.][]{aczel1966lectures, DiewertFunctional}. The non-trivial solution to this system of functional equations takes the form $f_j(x)=\alpha_jx^{\rho_{1j}}$ and $g_i(x)=\beta_ix^{\rho_{2i}}$ for any $\rho_{1j}, \rho_{2i} \in R$ and $f_j(1)=\alpha_j$ and $g_i(1)=\beta_i$. But note that\\\\
\hspace*{.5cm} $\delta=\dfrac{f_1(\gamma)}{f_1(1)}=....= \dfrac{f_M(\gamma)}{f_M(1)}=\dfrac{g_1(1/\gamma)}{g_1(1)}=....=\dfrac{g_N(1/\gamma)}{g_N(1)}\Rightarrow \gamma^{\rho_{1,1}}=....=\gamma^{\rho_{1,M}}=\gamma^{-\rho_{2,1}}=....=\gamma^{-\rho_{2,N}}$\\\\
But since this is true for every $\gamma>0$; we must have ${\rho_{1,1}}=....={\rho_{1,M}}={-\rho_{2,1}}=....={-\rho_{2,N}}=\rho$.\\\\
\noindent \textit{Proof of Sufficiency}: The proof is trivial.
\subsection{Existence of  GK, Generalized GK, EWGK and  the Generalized Mean Systems}
\noindent We can use Theorem-\ref{thm:th1} to establish the existence conditions for the GK system and its generalizations.
\textit{Corollary 1}: A necessary and sufficient condition for existence and uniqueness of the GK index (\ref{eq:3}), generalized GK (\ref{eq:4}), EWGK and the system based on generalized means (\ref{eq:5}) is connectedness of quantity matrix, $\mathbf{q}$.\\
\textit{Proof}: In system (\ref{eq:15}), defining $f_j(x)=\frac{1}{x}$, $g(x)=x$, $c_{ij}=p_{ij}q_{ij}$ and $d_{ij}=q_{ij}$ leads to (\ref{eq:3}). Existence and uniqueness of generalized GK index (\ref{eq:4}) can be proved by defining $f_j(x)=\frac{1}{x}$, $g(x)=x$, $c_{ij}=\beta_j p_{ij}q_{ij}$ and $d_{ij}=\beta_j q_{ij}$. \\
To prove the result for system (\ref{eq:5}), define
$f_j(x)=\dfrac{1}{x^\rho}$, $g(x)=x^\rho$, $c_{ij}=\beta_j p_{ij}q_{ij}$ and $d_{ij}=\beta_j q_{ij}$.\\ 
To prove the results for EWGK, define $f_j(x)=\frac{1}{x}$, $g(x)=x$, $c_{ij}=w_{ij}$ and $d_{ij}=w_{ij}/p_{ij}$. \\
\subsection{Existence Theorems for \cite{neary2004rationalizing} and \cite{rao1976existence} Systems} 
\noindent Theorem-\ref{thm:th1} does not cover all multilateral systems of interest. Theorem-\ref{thm:th3} below extends Theorem-\ref{thm:th1} to the case where the $d_{ij}$s and $c_{ij}$s are functions of $\mathbf{P}$. As we saw in Section \ref{sec:PPP_Multi}, \cite{rao1976existence} and \cite{neary2004rationalizing} 
 offer examples of such indexes. This specification makes the systems more complex and conditions for their existence and uniqueness more difficult to establish. \footnote{The theorem can be written in terms of functions of ${P}_i$s and ${PPP}_j$s with conditions stated in Theorem-\ref{thm:th2} but to avoid cumbersome notation here we focus on $f_j(x)=x$ and $g_i(x)=1/x$.}
\begin{thm} \label{thm:th3}
Consider the following general system of equations
\begin{subequations} \label{eq:18}
\begin{flalign}
\dfrac{1}{PPP_j}=\sum_{n=1}^{N}{\dfrac{d_{nj}(\mathbf{P,p,q})}{\sum_{n=1}^{N}{c_{nj}(\mathbf{P,p,q})}}\,P_n}\quad \quad \quad \quad \quad \quad  \quad \ (j=1,.....,M) \\ 
 P_i=\sum_{m=1}^{M}{\dfrac{c_{im}(\mathbf{P,p,q})}{\sum_{m=1}^{M}{d_{im}(\mathbf{P,p,q})}}\,\dfrac{1}{PPP_m} }\quad \quad \quad \quad \quad \quad \ \quad (i=1,.....,N)  
\end{flalign} 
\end{subequations}
Let $d_{ij}(\mathbf{P,p,q}) \geq 0$ and $c_{ij}(\mathbf{P,p,q}) \geq 0$ be continuous homogeneous functions of the same degree with respect to $\mathbf{P}$ then\\
\indent ($i$) there is at least one non-negative solution with some positive elements (up to a positive scalar multiple).\\
\indent ($ii$) there is at least one positive solution (up to a positive scalar multiple) if $d_{ij}(\mathbf{P,p,q})$ and $c_{ij}(\mathbf{P,p,q})$ are such that the vector function $\mathbf{G}$ defined below satisfies monotonicity and \hyperlink {SCF}{strong connectedness}.\\
\indent ($iii$) there is a unique positive solution (up to a positive scalar multiple) if $d_{ij}(\mathbf{P,p,q})$ and $c_{ij}(\mathbf{P,p,q})$ are such that the vector function $\mathbf{G}$  satisfies monotonicity and \hyperlink {Ind}{ indecomposibility}. 
\end{thm}
\noindent \textit{Proof}: Substitute $PPP_j$ from (\ref{eq:18}a) into (\ref{eq:18}b) and define $\mathbf{G}$ as
\begin {equation*}
 G_i=\sum_{m=1}^{M}{\dfrac{c_{im}(\mathbf{P,p,q})\sum\limits_{n=1}^{N}{\dfrac{d_{nm}(\mathbf{P,p,q})}{\sum_{n=1}^{N}{c_{nm}(\mathbf{P,p,q})}}\,P_n}}{\sum\limits_{m=1}^{M}{d_{im}(\mathbf{P,p,q})}}}\quad \quad \quad \quad \quad \quad \ \quad (i=1,.....,N)  
 \end{equation*} 
It is easy to see that $\mathbf{G}$ satisfies conditions (1) and (2) of \hyperlink{NET}{the nonlinear Perron-Frobenius theorem}\footnote{see \ref{sec:Toolkit} for detailed discussion of eigenvalue theorems.} and therefore the system $\mathbf{G(P)}=\lambda \mathbf{P}$ has a non-negative solution \citep[see e.g.][pp 151 for a simple proof]{nikaido2016convex}. If $\mathbf{G}$ also satisfies (3) and (4), there is at least one positive solution for the system and if $\mathbf{G}$ satisfies indecomposibility the positive solution is unique. 

The next step is to establish that the eigenvalue associated with the solution is equal to one (i.e. $\lambda^*=1$). Since there is at least one solution, we can write:
 \begin {equation*}
\lambda^*P_i^*=\dfrac{\sum\limits_{m=1}^{M}{c_{im}(\mathbf{P^*,p,q})\sum\limits_{n=1}^{N}{\dfrac{d_{nm}(\mathbf{P^*,p,q})}{\sum_{n=1}^{N}{c_{nm}(\mathbf{P^*,p,q})}}\,P_n^*}}}{\sum\limits_{m=1}^{M}{d_{im}(\mathbf{P^*,p,q})}}\quad \quad \quad \quad \quad \quad \ \quad (i=1,.....,N)  
 \end{equation*} 
or 
 \begin {equation*}
\lambda^*\bigg({\sum\limits_{m=1}^{M}{d_{im}(\mathbf{P^*,p,q})}}\bigg)P_i^*=\sum\limits_{m=1}^{M}{c_{im}(\mathbf{P^*,p,q})\sum\limits_{n=1}^{N}{\dfrac{d_{nm}(\mathbf{P^*,p,q})}{\sum_{n=1}^{N}{c_{nm}(\mathbf{P^*,p,q})}}\,P_n^*}}\quad \quad \quad \quad \quad \quad \ \quad (i=1,.....,N)  
 \end{equation*} 
Summing the equations over $i$ we have \footnote{Assuming that for each $j$ there is at least one $i$ for which $c_{ij}>0$ and for each $i$ there is at least one $j$ for which $d_{ij}>0$.}
\begin {align*}
\lambda^*\sum_{i=1}^{N}{\bigg({\sum\limits_{m=1}^{M}{d_{im}(\mathbf{P^*,p,q})}}P_i^*}\bigg)&=\sum\limits_{i=1}^{N}{\bigg(\sum\limits_{m=1}^{M}{c_{im}(\mathbf{P^*,p,q})\sum\limits_{n=1}^{N}{\dfrac{d_{nm}(\mathbf{P^*,p,q})}{\sum_{n=1}^{N}{c_{nm}(\mathbf{P^*,p,q})}}\,P_n^*}}\bigg)}\\
&=\sum\limits_{m=1}^{M}{\bigg(\dfrac{\sum\limits_{i=1}^{N}{c_{im}(\mathbf{P^*,p,q})}}{\sum_{n=1}^{N}{c_{nm}(\mathbf{P^*,p,q})}}\sum\limits_{n=1}^{N}{d_{nm}(\mathbf{P^*,p,q})\,P_n^*}\bigg)}\\
&=\sum\limits_{m=1}^{M}{\bigg(\sum\limits_{n=1}^{N}{d_{nm}(\mathbf{P^*,p,q})\,P_n^*}\bigg)}
 \end{align*} 
\begin{equation*}
\hspace*{4cm}\Rightarrow \lambda^*=1
\end{equation*}
This establishes the existence of a positive solution to the system.\\\\
Note that the system has at least one non-negative solution under very general conditions but in order to have a unique positive solution (using the nonlinear eigenvalue theorems in \ref{sec:Toolkit}), we need monotonicity and indecomposibility. Assuming monotonocity and $q_{ij}>0\Rightarrow d_{ij}>0 \Rightarrow c_{ij}>0$ using arguments similar to those used in proving Lemma-\ref{lm:lm2}, one can prove that $\mathbf{G}$ is indecomposable if $\mathbf{q}$ is connected but without knowing the exact form of $c_{ij}$ and $d_{ij}$s or placing stringent conditions on these functions, we cannot prove monotonicity of $\mathbf{G}$.\\
\textit{Corollary 2}: The Neary system (\ref{eq:10}), has at least one non-negative solution. It has a unique positive solution if $\mathbf{q}$ is connected and $\mathbf{G}$ satisfies monotonicity.\footnote{It may be difficult to show monotonicity for  \cite{neary2004rationalizing} as it includes $q^*$ and $q$ in the definitions.}\\
\textit{Proof}: In System (\ref{eq:18}) define $c_{ij}=p_{ij}q_{ij}$ and $d_{ij}=q_{ij}^*$ where it is evident that the $c_{ij}$ are homogenous of degree zero and the $d_{ij}$ are homogenous of degree zero with respect to $\mathbf{P}$ since the $q_{ij}^*$ are derived through (\ref{eq:10}c) \footnote{ The \cite{rao1976existence} system satisfies monotonicity but the weights are not defined in the form specified in Theorem-\ref{thm:th3} and it is not possible to prove $\lambda^*=1$. }.

\subsection{Existence Theorems for \cite{rao1990system}, IDB and Related Systems}
\noindent We now focus on a variant of Theorem-\ref{thm:th1} where the system cannot be turned into a linear system in terms of $g_i$ and $f_j$. This theorem can be used to establish existence and uniqueness of solutions for systems that are not covered by system (\ref{eq:15}) such as the multilateral index numbers proposed in \cite{rao1990system}, IDB, \footnote{The existence of IDB system has been proved by \citet{balk1996comparison} and \cite{diewert2013methods}.}  and the arithmetic index (\ref{eq:8}) described in Section \ref{sec:PPP_Multi}.

\begin{thm} \label{thm:th4}
Suppose $f_j$ and $g_i$ are known non-negative, bijective functions defined over non-negative values and
\begin{subequations} \label{eq:19}
\begin{flalign}
f_j(PPP_j)=\sum_{n=1}^{N}{\dfrac{a_{nj}c_n}{g_n(P_n)}}\quad \quad \quad \quad \quad \quad  \quad \ (j=1,.....,M) \\ 
 g_i({P_i})=\sum_{m=1}^{M}{\dfrac{a_{im}d_m}{f_m(PPP_m)}}\quad \quad \quad \quad \quad \quad \ \quad (i=1,.....,N)  
\end{flalign} 
\end{subequations}
with $a_{ij} \geq 0$, $c_i>0$ and $d_j>0$, then \hyperlink{COM}{compatibility} and \hyperlink{CONN}{connectedness} conditions (as defined in \ref{sec:Toolkit}) are necessary and sufficient for having a unique positive solution. 
\end{thm}
\noindent \textit{Proof}: Substituting (\ref{eq:19}.b) into (\ref{eq:19}.a), we obtain
\[
f_j=\sum_{n=1}^{N}{\dfrac{a_{nj}c_n}{\sum\limits_{m=1}^{M}{\dfrac{a_{nm}d_m}{f_m}}}}\quad \quad  (j=1,....,M)
\]
This is exactly the \hyperlink{DAD}{DAD Problem} discussed in \ref{sec:Toolkit} where compatibility and connectedness conditions are necessary and sufficient for existence of a unique solution. 

In this general form, we cannot relate the solution of the system to connectedness of the countries. However, there is a more interesting special case of this system where we can show that the connectedness of $\mathbf{q}$ is both necessary and sufficient for existence of a unique positive solution. This system covers the systems defined in equations (\ref{eq:6}), (\ref{eq:7}), (\ref{eq:8}) and (\ref{eq:9}) as special cases.

\begin{thm} \label{thm:th5}
Consider the following system of equations:
\begin{subequations} \label{eq:20}
\begin{flalign}
f_j=\sum_{n=1}^{N}{\dfrac{d_{nj}}{\sum_{n=1}^{N}d_{nj}}\dfrac{e_{nj}}{g_n}}\quad \quad \quad \quad \quad \quad  \quad \ (j=1,.....,M) \\ 
 g_i=\sum_{m=1}^{M}{\dfrac{d_{im}}{\sum_{j=m}^{M}d_{im}}\dfrac{e_{im}}{f_m}}\quad \quad \quad \quad \quad \quad \ \quad (i=1,.....,N)  
\end{flalign} 
\end{subequations}
where $d_{ij}>0 \Leftrightarrow e_{ij}>0 \Leftrightarrow q_{ij}>0$. Then connectedness of countries through quantity matrix $\mathbf{q}$ is both necessary and sufficient for uniqueness of the solutions (up to a positive scalar multiple).
\end{thm}
\noindent \textit{Proof}: Note first that this system is a special case of (\ref{eq:19}) by defining $a_{ij}=\dfrac{d_{ij}e_{ij}}{(\sum_{m=1}^{M}{d_{im}})(\sum_{n=1}^{N}{d_{nj}})}$, $c_i=\sum_{j=1}^{M}{d_{ij}}$ and $d_j=\sum_{i=1}^{N}{d_{ij}}$. \\\\
\textit{Sufficiency}: Define $\mathbf{D}=\{d_{ij}\,|\, i=1,...,N, j=1,...,M\}$, $\mathbf{A}=\{a_{ij} | i=1,...,N, j=1,...,M\}$ and let $I\subset \{1,...,N\}, J\subset\{1,...,M\}$ with $\mathbf{A}_{I^cJ^c}=\mathbf{0}$ where $I^c$ is complement of $I$ and $\mathbf{A}_{IJ}=\{a_{ij}| i\in I \, \text{and} \, j \in J\}$. Since $d_{ij}>0 \Leftrightarrow e_{ij}>0 \Leftrightarrow q_{ij}>0$, connectedness of $\mathbf{q}$ is equivalent to connectedness of $\mathbf{A}$ and $\mathbf{D}$. To show that compatibility is automatically satisfied note that
\begin{equation} \label{eq:21}
\sum_{i \in I^c}{}{c_i}=\sum_{i \in I^c}{}{\sum_{j \in J^c\cup J}{}{d_{ij}}}=\sum_{i \in I^c}{}{\sum_{j \in J}{}{d_{ij}}} \,\,\, \text{since} \sum_{i \in I^c}{}{\sum_{j \in J^c}{}{d_{ij}}}=0
\end{equation}
but we also have
\begin{equation} \label{eq:22}
\sum_{j \in J}{}{d_j}=\sum_{i \in I^c\cup I}{}{\sum_{j \in J}{}{d_{ij}}}=\sum_{i \in I^c}{}{\sum_{j \in J}{}{d_{ij}}}+\sum_{i \in I}{}{\sum_{j \in J}{}{d_{ij}}}\geq
\sum_{i \in I^c}{}{\sum_{j \in J}{}{d_{ij}}}
\end{equation}
\[\hspace*{2cm}\text{(\ref{eq:21}) and (\ref{eq:22})}\Leftrightarrow \sum_{j \in J}{}{d_j} \geq \sum_{i \in I^c}{}{c_i}\] 
It is also obvious that the inequality is strict if and only if $\mathbf{A}_{IJ} \neq \mathbf{0}$ (or $ \mathbf{D}_{IJ}\neq \mathbf{0}$) which proves compatibility. To illustrate the above argument, consider the following decomposition of $\mathbf{D}$
after appropriate re-ordering of rows and columns:\\
\begin{figure}[h]
\centering
\includegraphics[scale=1]{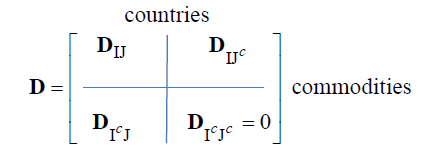}
\end{figure}

\noindent Since $\mathbf{D}_{I^cJ^c}=\mathbf{0}$, the sum of the elements in $\mathbf{D}_{I^cJ}$ i.e. $\sum_{i \in I^c}{}{\sum_{j \in J}{}{d_{ij}}}$ is equal to $\sum_{i \in I^c}{}{c_i}$, but $\sum_{j \in J}{}{d_j}$ is sum of the elements in $\mathbf{D}_{I^cJ}$ and $\mathbf{D}_{IJ}$.\\
\textit{Necessity}: According to the DAD theorem, a unique positive solution to (\ref{eq:20}) $\Leftrightarrow $ compatibility and connectedness of $\mathbf{A}=\{a_{ij} | i=1,...,N, j=1,...,M\}$  $\Leftrightarrow$ connectedness of $\mathbf{D} \Leftrightarrow$ connectedness of $\mathbf{q}$ due to the assumption $d_{ij}>0 \Leftrightarrow q_{ij}>0$ .\\

If the condition that, for every $\gamma>0$ both $[\mathbf{PPP^* ,P^*}]$ and $[\gamma \mathbf{PPP}^*,(1/\gamma)\mathbf{P}^*]$ are solutions to the system (\ref{eq:19}), is imposed then using arguments similar to those in Theorem-\ref{thm:th3} above, the functional forms used in the system must be set as $f_j(x)=\alpha_j x^\rho$ and $g_i(x)=\beta_i x^\rho$ for any
$\alpha_j>0$, $\beta_i>0$ and $\rho \in R$.\\
\noindent \textit{Corollary 3}: If matrix $\mathbf{q}$ is connected, systems (\ref{eq:6}), (\ref{eq:7}), (\ref{eq:8}) and (\ref{eq:9}) have unique positive solutions.
We consider system (\ref{eq:12}) which encompasses all the others. Note that if in (\ref{eq:20}) we define $f_j(x)=g_i(x)=x^\rho$ and $d_{ij}=w_{ij}$, $e_{ij}=p_{ij}^\rho$, $c_i=\sum_{j=1}^{M}{w_{ij}}$ and $d_j=\sum_{i=1}^{N}{w_{ij}}=1$ we obtain system (\ref{eq:12}) and therefore Theorem 5 proves its existence and uniqueness.
\section{Conclusion}\label{sec:conclusions}
\noindent This paper provides general theorems for establishing existence and uniqueness of positive solutions to multilateral index number systems that make use of the twin concepts of international average prices and purchasing power parities in the context of international comparisons of prices and real incomes. The main result indicates that connectedness of the matrix of quantities or equivalently connectedness of the associated quantity-adjacent graph is in general both necessary and sufficient for existence of a unique positive solution. The theorems proved in the paper are general and powerful enough to prove the existence and uniqueness of solutions not only to the currently used system of index numbers but also systems that may come into vogue in the future. While simple connectedness guarantees the existence of solutions and therefore the viability of most of the multilateral index number systems, the strength of connectedness could have implications for the reliability of the indexes which are being studied in our other works.

\bibliographystyle{model5-names}
\bibliography{bbib-PPP}

\appendix
\section{Mathematical Toolkit for Existence Theorems}\label{sec:Toolkit}
\subsection{Connectedness, Graphs and Irreducibility}
\noindent In this section, we present the notions of connectedness across countries, (strong) connectedness of graphs, and irreducibility and their relationships to each other.\\
\hypertarget {CM} {\textbf{\textit{Connectedness of Matrix $\mathbf{q}$}}: The quantity matrix $\mathbf{q}=\{q_{ij}\}$, an $N\times M$ matrix of quantities where $q_{ij}$ is the quantity of  $i$-th commodity consumed in  $j$-th country, is said to be connected if the set of all countries cannot be split into two or more disjoint subsets such that there are no commodities that are commonly consumed across the two groups.}\\
More formally, a non-negative matrix of quantities $\mathbf{q}$ is said to be connected if for every nonempty proper subset of countries $J\subset \{1,2,...,M\} \neq \varnothing$ there exists at least one country $j\in J$, one country $l\notin J$ and one commodity $k \in {1,...., N}$ such that both $q_{kj}>0$ \, and \, $q_{kl}>0$.  An equivalent mathematical definition of connectedness is that for every non-empty proper subset $I\subset \{1,2,...,N\}$ and $J\subset \{1,2,...,M\}$, with $I^c$ and $J^c$ as their complements, $\mathbf{q}_{I^cJ^c}=0$ implies $\mathbf{q}_{IJ} \neq 0$ where $\mathbf{q}_{IJ}=\{{q}_{ij}, i \in I\, \text{and}\, j \in J\}$.\\
Connectedness of matrix $\mathbf{q}$ is critical in international comparisons. If $\mathbf{q}$ is disconnected, then countries can be divided into two groups with no common items of consumption. In such a case, there is no basis for making price comparisons. Thus, connectedness of $\mathbf{q}$ seems necessary. In this paper, we show that this condition is both necessary and sufficient for uniqueness of several classes of multilateral index number systems. It is possible to give this notion of connectedness a graph theoretic interpretation.\\
\textbf{\textit{Quantity-Adjacent Graph}}: Let $\mathbf{\mathcal{G}}_q$ represent a graph associated with a given quantity matrix $\mathbf{q}$ with countries as vertices of the graph. Two vertices $j$ and $k$ are connected by an edge if there exists a commodity $i\in \{1,2,...,N\}$ such that $q_{ij}>0$ and $q_{ik}>0$.\\
\hypertarget {CA}{\textbf{\textit{Connected Graph}}}: A graph, $\mathbf{\mathcal{G}}_q$ is said to be connected if for any pair of countries $j$ and $k$, there exists a sequence of countries  $\{j_1,...,j_k\}\in\{1,2,...,\}$ such that each consecutive pair of countries in the sequence are connected by an edge.\\
It has been established that the graph $\mathbf{\mathcal{G}}_q$ associated with a quantity matrix $\mathbf{q}$ is connected if and only if the quantity matrix $\mathbf{q}$ is connected .\\
\textbf{\textit{Connected Matrices and Adjacent Graphs-An illustration}}: The notions of connectedness of the matrix of quantities and the connectedness of the graphs associated with quantity matrices are central to the existence theorems proved in the paper. We illustrate these notions using three examples where we consider four countries and four commodities. The quantity matrix, $\mathbf{q}$, in this case is $4\times 4$ where columns represent countries and the rows represent commodities.\\
The first example is where all the four commodities are consumed in all the countries. As a result, the graph associated with the matrix will have edges connecting all pairs of countries as shown below. 
\begin{figure}[h]
\centering
        \begin{tabular}{p{5cm}c}
            {$\displaystyle
               \mathbf{q} =                 {\renewcommand{\arraystretch}{1.2}
                \begin{pmatrix}
   10 & 5 & 100 & 25  \\
   6 & 3 & 75 & 35  \\
   80 & 25 & 250 & 125  \\
   8 & 6 & 35 & 40  \\
                \end{pmatrix}}
            $}
            &       $\vcenter{\hbox{\includegraphics[scale=.7]{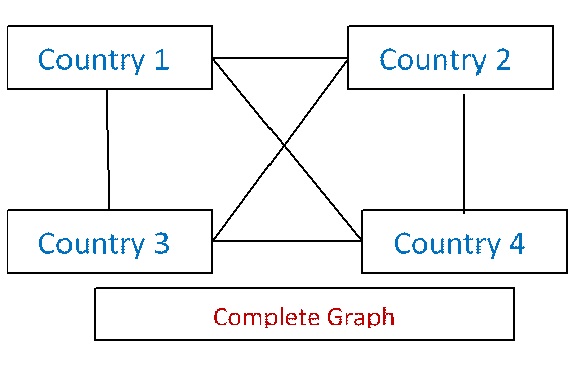}}}$
        \end{tabular}
\end{figure}

\noindent In this case, it is easy to see that the matrix $\mathbf{q}$ is connected. The adjacent graph is a complete graph where each country is directly connected with every other country. This is a case where there is complete connectedness.

The second example is where the first country consumes all four commodities whereas the other three countries consume only one or two of the commodities. The quantity matrix and the adjacent graphs in this case are of the following form.

\begin{figure}[h]
\centering
        \begin{tabular}{p{5cm}c}
            {$\displaystyle
               \mathbf{q} =                 {\renewcommand{\arraystretch}{1.2}
                \begin{pmatrix}
   10 & 5 & 0 & 0  \\
   6 & 0 & 75 & 0  \\
   80 & 25 & 0 & 0  \\
   8 & 0 & 0 & 40  \\
                \end{pmatrix}}
            $}
            &       $\vcenter{\hbox{\includegraphics[scale=.65]{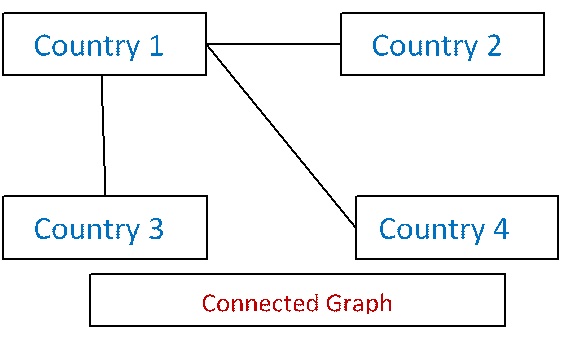}}}$
        \end{tabular}
\end{figure}
\noindent It is easy to check that the matrix $\mathbf{q}$ here is connected. We can see that Country 1 is connected to the rest of the countries. However, there are no direct links between Countries 2, 3 and 4. In this case the graph is connected with no cycles – this means between any two countries there is
only one chained path connecting the two countries. For example, Countries 3 and 2 are connected through Country 1. This type of graph is referred to as a spanning tree. Here there is connectivity only through Country 1 but it is still sufficient for connectedness of the matrix $\mathbf{q}$. \\
Finally, we consider an example where the four countries are divided into two separate groups, Countries 1 \& 2 and Countries 3 \& 4. These groups have no commodities commonly consumed. Therefore, the matrix $\mathbf{q}$ is not a connected matrix and it consists of two sets of matrices of lower dimensions (two) each of which are connected but the full matrix is not connected.

\begin{figure}[h]
\centering
        \begin{tabular}{p{5cm}c}
            {$\displaystyle
               \mathbf{q} =                 {\renewcommand{\arraystretch}{1.2}
                \begin{pmatrix}
   10 & 5 & 0 & 0  \\
   6 & 3 & 0 & 0  \\
   0 & 0 & 250 & 125  \\
   0 & 0 & 35 & 40  \\
                \end{pmatrix}}
            $}
            &       $\vcenter{\hbox{\includegraphics[scale=.7]{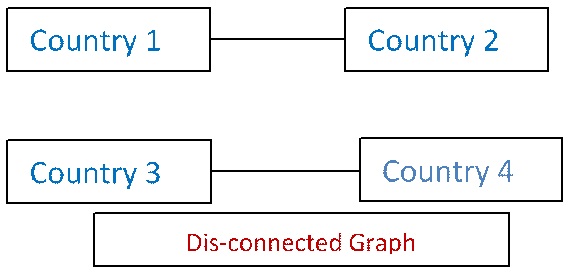}}}$
        \end{tabular}
\end{figure}
\noindent In this example, countries 1 \& 2 have commodities (1 and 2) commonly consumed whereas countries 3 \& 4 consume only commodities 3 and 4. In this case, the adjacent graph is disconnected. Countries 3 \&4 have no connections with Countries 1\&2. In this case there is no basis for multilateral comparisons involving all four countries.

There are well-known algorithms such as the Breadth-First Search (BFS) which can be used for checking connectedness of a graph (and as a result $\mathbf{q}$). There are also very simple sufficient conditions that guarantee  connectedness: ($i$) there is one commodity that every country consumes; ($ii$) there is one country that consumes all commodities. These conditions have been used by e.g. \citet{balk2008price} in his proofs of existence.\\
We now introduce several other definitions that are closely related to the connectedness of $\mathbf{q}$ and the existence of multilateral index numbers.\\
\hypertarget {IR} {\textbf{\textit{Irreducibility of a Square Matrix}}: A non-negative $M\times M$ matrix $\mathbf{A}=\{{{a}_{ij}}\}$ is said to be irreducible if and only if for any proper subset $J\subset \{1,2,...,M\}\ne \varnothing $ there exists at least one $j\in J$ and $i\notin J$ such that ${{a}_{ij}}>0$.}\\
\textbf{\textit{Strong Connectedness of the Directed Graph Associated with Matrix A}}:  
Let the directed graph (digraph) associated to matrix $\mathbf{A}$, $\Theta(\mathbf{A})$ be a directed graph (digraph) with vertices of $1,...,M$ and an edge from $i$ to $j$ if and only if ${{a}_{ij}}>0$. This digraph is said to be strongly connected if for any pair of vertices of $\Theta(\mathbf{A})$, there is a path that connects them.
It is a well-known result that irreducibility of a matrix and strong connectedness of its digraph are equivalent. The following definition generalizes the strong connectedness to nonlinear functions.\\
\hypertarget {SCF} {\textbf{\textit{Strong Connectedness of Digraph Associated with a Function G}}}: For a function $\mathbf{G(x)}:R_{++}^{M}\to R_{++}^{M}$ , let $\Theta(\mathbf{G})$ be the digraph with vertices of $1,...,M$ and an edge from $i$ to $j$ if and only if $\underset{x_{j}\to \infty }{\mathop{\lim }}\,\ {{G}_{i}}({{\mathbf{x}}_{\{j\}}})=\infty $ (where vector ${{\mathbf{x}}_{\{j\}}}$ refers to a vector that has all elements equal to one except for its $j$th element). Then $\mathbf{G}$  is strongly connected if for any pair of vertices, there is a path that connects them. It is easy to see that when $\mathbf{G}(\mathbf{x})$ is linear (i.e. $\mathbf{G}(\mathbf{x})=\mathbf{Ax}$), this is equivalent to strong connectedness of matrix $\mathbf{A}$.\footnote{See Gaubert and Gunawardena (2004) for further information on strong connectedness, more examples and its relevance to nonlinear eigenvalue theorems.}
As an example, consider the following nonlinear function
\begin{equation*} \hspace*{2cm}
\mathbf{G(x)}=\left(\begin{matrix} \min \left\{ \sqrt{x_1x_2},2\sqrt{x_2{x_3}} \right\} \\ 
 \max \left\{ \sqrt{x_2x_3},3\sqrt{x_3x_1} \right\} \\ 
  \max \left\{ {x_1},{x_3} \right\} \end{matrix}\right) 
\end{equation*}
 The digraph associated to $\mathbf{G}$ is strongly connected as the following figure (constructed based on the above definition) shows
\begin{figure}[h]
\centering
\includegraphics[scale=1]{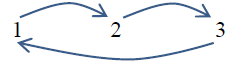}
\end{figure}

\noindent \hypertarget{Ind}{\textbf{\textit{{Indecomposibility of a Function}}}}: For any pair of vectors $\mathbf{x}$ and $\mathbf{y}$ with $\mathbf{x} \geq \mathbf{y}\geq \mathbf{0}$ define a nonempty proper subset $\Omega=\{j\,|\, x_j>y_j\} \subset {1,...,M}$. Function $\mathbf{G}$ is indecomposable if there exists $k \notin \Omega$  such that $G_k(x_1,...x_M) \neq G_k(y_1,...y_M)$. This concept of indecomposibility is equivalent to the irreducibility of a matrix if $\mathbf{G}$ is linear. Indecomposibility is generally a stronger condition than strong connectedness.
\subsection{Nonlinear Perron-Frobenius Theorems} \label{sec:A2}
\noindent We start with a general multilateral system of $M+N$ equations of the following form:
\begin{align*}
  P_i=H_{i}^{1}(\mathbf{PPP,p,q})\quad \quad \quad \quad \quad \quad \quad \quad \quad (i=1,.....,N)   \\ 
  PPP_j=H_{j}^{2}(\mathbf{P,p,q})\quad \quad \quad \quad \quad \quad \quad \quad  \ \ (j=1,.....,M) 
\end{align*}
where $\mathbf{PPP}$ and $\mathbf{P}$ are respectively $M \times 1$ and $N \times 1$ vectors of purchasing power parities and world average prices. We can substitute equation $P_i$ into the equation for $PPP_{j}$ leading to a homogeneous system of $M$ equations of the form:
\begin{flalign} \label{eq:12}
 PPP_j=H_{j}^{2}(H_{i}^{1}(\mathbf{PPP,p,q}),\mathbf{p,q}) \quad \quad \quad \quad \quad \quad \quad \quad \quad \quad  \\
 =G_{j}(\mathbf{PPP,p,q})\quad \quad \quad \quad \quad \quad \quad \quad \ \ (j=1,.....,M) \notag
\end{flalign}
This means that the set of equations in (\ref{eq:12}) are such that each $PPP_j$ is expressed as a function of observed price and quantity data as well as all $PPP_j$s. Solving equations in (\ref{eq:12}) is equivalent to solving the following general system of homogeneous equations:
\begin{equation} \label{eq:13}
\hspace*{0.5cm} x_j=G_j(x_1,x_2,...,x_M)\,\,\,\,\,\,j=1,2,...,M
\hspace*{0.5cm} \text{or in vector form} \hspace*{0.5cm} \mathbf{x}=\mathbf{G}(\mathbf{x})
\end{equation}
To prove the existence of solutions to (\ref{eq:12}) or (\ref{eq:13}), Perron-Frobenius (or eigenvalue) theorems are the basic tools. There are a host of such theorems which give conditions for existence of a solution under various situations \citep[see e.g. the book by][for a review of eigenvalue theorems]{lemmens2012nonlinear}. The following theorem is one of the general eigenvalue theorems due to \cite{gaubert2004perron} and \cite{morishima1964equilibrium}. \\\\
\hypertarget {NET} {\textbf{\textit{Nonlinear Perron-Frobenius Theorem}}}: Let $G_j(x_1,x_2,...,x_M)$ for $j=1,...,M$ satisfy the following conditions:
 \begin{enumerate} 
\item $\mathbf{G}(\mathbf{x})$ is a function $R_{++}^{M}\to R_{++}^{M}$
\item Homogeneity: functions $G_j(\mathbf{x})$ for $j=1,...,M$ are homogeneous of degree one in $\mathbf{x}$
\item Monotonicity: for all $\mathbf{x}\ge \mathbf{y}$,  $G_j(\mathbf{x})\ge G_j(\mathbf{y})$ for $j=1,....,M$
\item Strong connectedness of the digraph associated with $\mathbf{G}(\mathbf{x})$ 
\label{item:4}
\item Indecomposibility of function $\mathbf{G}(\mathbf{x})$ 
\label{item:5}
\end{enumerate}
and consider the system of equations
\begin{equation} \label{eq:14}
\hspace*{2cm} G_j(x_{1},.....,x_{M})=\lambda x_{j} \quad \quad \quad \quad \quad j=1,....,M       
\end{equation}
(i) under conditions (1) and (2), there is at least one non-negative $\mathbf{x}^*$(up to a positive scalar \hspace*{0.5cm} multiple) and $\lambda^*$ that satisfy the equation.\\
(ii) under conditions (1),(2), (3) and (4), there is at least one positive $\mathbf{x}^*$(up to a positive scalar \hspace*{0.5cm} multiple) and $\lambda^*$ that satisfy the equation.\\
(iii) under conditions (1),(2), (3) and (5), there is a unique positive $\mathbf{x}^*$(up to a positive scalar \hspace*{0.5cm} multiple) and $\lambda^*$ that satisfy the equation.\\

A few remarks on the relevance of nonlinear eigenvalue theorems are:
\begin{enumerate} [\textbullet]
\item Eigenvalue theorems play an important role in establishing conditions for the existence of solutions to the multilateral index number systems examined in this paper. It is sufficient if the two functions, $H_{i}^{1}$and $H_{j}^{2}$, are such that the function $\mathbf{G}$ obtained through (\ref{eq:12}) satisfies the conditions stated in the nonlinear eigenvalue theorems.
\item In the context of multilateral systems, we need to find a solution that satisfies equation (\ref{eq:12}), i.e. $\mathbf{x}\,\,\mathbf{=}\,\,\mathbf{G}(\mathbf{x})$. For this to hold, we need to also establish that $\lambda^*=1$ in equation (\ref{eq:14}). To have this, we often need some extra conditions on $H_{i}^{1}$ and $H_{j}^{2}$ which we may refer to as \textit{compatibility} conditions.
\item In all eigenvalue theorems, it is assumed that the function $\mathbf{G}(\mathbf{x})$ is continuous, homogenous and monotone, conditions which are easy to check but which are often not sufficient for a positive solution. A further assumption such as \ref{item:4} or \ref{item:5} above (which could be different across eigenvalue theorems) is often required. In this paper, we provide easily verifiable conditions to check such conditions in terms of the quantity data matrix.
\item \hypertarget{MNET} {A well-known eigenvalue theorem} in mathematical economics proved by \citet[pp 195-199]{morishima1964equilibrium} and \citet[pp 149-161]{nikaido2016convex} gives a sufficient condition for existence of a unique positive solution to the above problem by replacing assumption (\ref{item:4}) above with (\ref{item:5}) i.e. \textit{indecomposibility} of function $\mathbf{G}(\mathbf{x})$. The linear version of this theorem (i.e. $\mathbf{G}(\mathbf{x})=\mathbf{Ax}$) is the well-known Perron-Frobenius theorem \citep[][Chapter 3]{gantmacher2005applications} where the required sufficient condition for a unique positive solution is irreducibility of the non-negative matrix $\mathbf{A}$.
\end{enumerate}
To prove existence of some of the multilateral systems, we need to appeal to another nonlinear eigenvalue theorem known as the \textit{DAD theorem} \citep[see e.g. Chapter 7 of][or \cite{menon1969spectrum}]{lemmens2012nonlinear}. The DAD theorem entails a triplet $\{\mathbf{A},\mathbf{c},\mathbf{d}\}$ where $\mathbf{A}=\{{{a}_{ij}}\}$ is an $N\times M$ non-negative matrix, $\mathbf{c}$ is an $N\times 1$ positive vector and $\mathbf{d}$ is an $M\times 1$ positive vector. The DAD theorem, in its original form, gives conditions under which it is possible to find positive diagonal matrices 
$\mathbf{D}_{1}=diag\{[{{\delta }_{1}},...,{{\delta }_{N}}]\}$ and ${{\mathbf{D}}_{2}}=diag\{[{{\partial }_{1}},...,{{\partial }_{M}}]\}$ with ${{\mathbf{D}}_{1}}\mathbf{A}{{\mathbf{D}}_{2}}$ having row sums equal to $\mathbf{d}$ and column sums equal to $\mathbf{c}$. Note that
\[\hspace*{2cm} {{\mathbf{D}}_{1}}\mathbf{A}{{\mathbf{D}}_{2}}=\left[ \begin{matrix}
   {{\delta }_{1}}{{a}_{11}}{{\partial }_{1}} & \cdots  & {{\delta }_{1}}{{a}_{1M}}{{\partial }_{M}}  \\
   \vdots  & \ddots  & \vdots   \\
   {{\delta }_{N}}{{a}_{N1}}{{\partial }_{1}} & \cdots  & {{\delta }_{N}}{{a}_{NM}}{{\partial }_{M}}  \\
\end{matrix} \right]\]
Since columns sum to $\mathbf{c}$ and rows sums to $\mathbf{d}$, we must have 
\begin{alignat*}{2}
  & \begin{aligned} & \begin{cases}
  {{\delta }_{1}}\sum\limits_{m=1}^{M}{{{a}_{1m}}{{\partial }_{m}}={{c}_{1}}} \\ 
 \qquad \vdots \qquad \vdots  \\ 
 {{\delta }_{N}}\sum\limits_{m=1}^{M}{{{a}_{Nm}}{{\partial }_{m}}={{c}_{N}}}
 \end{cases}\\
   \end{aligned}
    & \& \quad &
  \begin{aligned} 
 {{\partial }_{1}}\sum\limits_{n=1}^{N}{{{a}_{n1}}{{\delta }_{n}}={{d}_{1}}} \\ 
\vdots \qquad \vdots \qquad \\
 {{\partial }_{M}}\sum\limits_{n=1}^{N}{{{a}_{nM}}{{\delta }_{n}}={{d}_{M}}} \\ 
\end{aligned}
\end{alignat*}
This is a system of $M+N$ equations in $M+N$ unknowns. Substituting ${\delta }_{i}$s in the set of equations on the right from equations on the left and defining ${{x}_{j}}={{{d}_{j}}}/{{{\partial }_{j}}}\;$, it is easy to see that this problem is equivalent to the following eigenvalue problem:\\
\hypertarget{DAD}{\textbf{\textit{DAD Eigenvalue Theorem}}}: Consider the following system of equations:              
\[\sum\limits_{n=1}^{N}{\dfrac{a_{nj}c_n}{\sum\limits_{m=1}^{M}{\dfrac{a_{nm}d_m}{x_m}}}}=x_j\quad  \text{for}\,\,\, j=1,....,M  \quad \quad  or \quad \quad  \mathbf{{A}'}\,\,\dfrac{\mathbf{c}}{\mathbf{A}\left( \dfrac{\mathbf{d}}{\mathbf{x}} \right)}=\mathbf{x}\]
where in the equation on the right, vector divisions are element by element.
Then a necessary and sufficient condition for existence of at least one positive solution is the \textit{compatibility condition} (defined below). A further condition of connectedness of $\mathbf{A}$ provides a necessary and sufficient condition for uniqueness.\\
\hypertarget{COM}{\textit{Compatibility Condition}}: For every $I\subseteq \{1,...,N\}$ and $J\subseteq \{1,...,M\}$ define $I^{c}$ and $J^{c}$ as complements of these sets. The compatibility condition implies that for every ${{\mathbf{A}}_{{{\text{I}}^{\text{c}}}{{\text{J}}^{\text{c}}}}}=\mathbf{0}$, the inequality $\sum\limits_{i\in {{\text{I}}^{\text{c}}}}{{{c}_{i}}}\le \sum\limits_{j\in \text{J}}{{{d}_{j}}}$ holds, and the inequality is strict if and only if ${{\mathbf{A}}_{\text{IJ}}}\ne \mathbf{0}$.\\ 
\hypertarget{CONN}{\textit{Connectedness  of $\mathbf{A}$}}:  For every $I\subset \{1,...,N\}$, $J\subset \{1,...,M\}$, ${{\mathbf{A}}_{{{\text{I}}^{\text{c}}}{{\text{J}}^{\text{c}}}}}=\mathbf{0}\Rightarrow {{\mathbf{A}}_{\text{IJ}}}\ne \mathbf{0}$. \\
To see why compatibility is required, note that, with  ${{\mathbf{A}}_{{{\text{I}}^{\text{c}}}{{\text{J}}^{\text{c}}}}}=\mathbf{0}$, after appropriate row and column permutations $\mathbf{D_1AD_2}$ can be written as 
\begin{figure}[h]
\centering
\includegraphics[scale=.75]{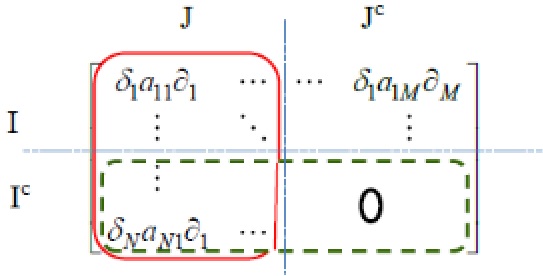}
\end{figure}

\noindent where $\sum\limits_{i\in I^c}{c_i}$ is the sum of the elements in the horizontal dashed rectangle and $\sum\limits_{j\in J}{d_j}$ is the sum of elements in the red vertical rectangle. Excluding the zeros, the horizontal dashed rectangle is a subset of the red vertical rectangle. Therefore, in order to have a compatible set of equations we require the compatibility condition. Connectedness is required for uniqueness because otherwise the system can be decomposed into two unrelated subsystems.
\subsection{Relationships between the Concepts}
 \noindent Assuming that the function $\mathbf{G}$ satisfies conditions (1) to (3) stated in the nonlinear eigenvalue theorem, the link between various concepts and eigenvalue theorems can be summarized as follows:
\begin{enumerate}[\textbullet]
\item Irreducibility of a square matrix (equivalently, strong connectedness of its digraph) is a sufficient condition for existence of a unique positive solution to a linear eigenvalue problem (Perron-Frobenius theorem).
\item Strong connectedness of the digraph of a function is a sufficient condition for existence of at least one solution to the nonlinear eigenvalue problem.
\item Indecomposibility is a sufficient condition for existence of a unique positive solution for the nonlinear eigenvalue problem.
\item Indecopmposibility is a stronger condition than strong connectedness but they are both equivalent to irreducibility if the function is linear.
\item Compatibility and connectedness of matrix $\mathbf{A}$ together provide both necessary and sufficient conditions for the DAD problem to have a unique positive solution.
\end{enumerate}

\end{document}